# Before Bethe: Early Ideas of the Sun's Generation of Energy

Helge Kragh[*]

**Abstract:** Solar energy remained an enigma for nearly a century. For a while astronomers and physicists believed that the source of the Sun's energy was gravitational contraction, but the theory turned out to be untenable. Inspired by the new science of radioactivity, by the early twentieth century they increasingly focused on subatomic processes. The paper outlines the attempts to solve the solar energy problem in the period from about 1905 to 1938, which in the later phase involved nuclear astrophysics, a new and exciting interdisciplinary science. Although the problem was only satisfactorily solved with Hans Bethe's theory of the late 1930s, the earlier attempts involving scientists such as Arthur Eddington, Robert Atkinson and George Gamow can be considered a slow preparation for Bethe's breakthrough.

## 1. Introduction

Referring to the "great mystery" of how the Sun maintains its vast emission of heat and light, in his *Treatise on Astronomy* from 1833 the British astronomer and natural philosopher John Herschel despairingly wrote: "Every discovery in chymical science here leaves us completely at a loss, or rather, seem to remove further the prospect of a probable explanation." Whatever the Sun was, it was not a chemical machine powered by coal or other combustible materials. Herschel (1833, 202) concluded that "we know nothing, and we may conjecture everything."

Five years after Herschel admitted his complete ignorance, the French physicist Claude Pouillet, director of the Conservatoire des Arts et Métier in Paris, succeeded in measuring the heating effect of sunlight by means of a specially designed calorimeter. Taking into account the decrease in the intensity of sunlight caused by the atmosphere, Pouillet reported the value of the "solar constant" – a name he coined – to be 1.76 calories per square centimetre per minute (Kidwell 1981). In modern units it corresponds to 1.23 kW/m² and a total solar power output of the order $10^{26}$ W. Pouillet's values were close to those accepted today, namely 1.36 kW/m² for the solar constant and $3.85 \times 10^{26}$ W for the power.

The recognition of the Sun's enormous output of heat naturally provoked the question of how the heat was generated, a question which could only be addressed on a scientific basis after the law of energy conservation was established a few years later. One thing was to address the question, another was to answer it satisfactorily, which was first done by the German-American physicist Hans Bethe in a couple of celebrated papers from the late 1930s. With Bethe's theory Herschel's "we know nothing" was reverted to "we know a lot." The present paper does not deal specifically with Bethe's work but focuses on earlier and less well-known theories of solar energy as caused by subatomic processes. After a decade of speculations in the early part of the twentieth century on radioactivity as a possible source of the Sun's energy, during the 1920s a few astronomers and physicists suggested nuclear reactions as an alternative. However, it was only with quantum mechanics and the proton-neutron picture of the atomic nucleus that nuclear astrophysics took off and eventually led to an understanding of the basic mechanism of solar energy (Hufbauer 2006; Kragh 2021, 221-252).

## 2. From gravitation to radioactivity

During the last third of the nineteenth century many astronomers and physicists were confident that what had so puzzled John Herschel was no longer an unsolved mystery. The answer to the question of solar heat generation, they thought, was that the Sun contracted slightly and that the gravitational energy released during the process manifested itself in the liberation of light and heat. The gravitational contraction theory was first proposed by Hermann Helmholtz in an important address of 1854 and further developed by William Thomson in a series of works from about 1860 to 1900. What became known as the Thomson-Helmholtz theory came to be celebrated as the authoritative and probably correct theory of solar energy production during the late Victorian period (Kragh 2016).[1] And yet, although the theory was generally accepted and for half a century had the status of a standard model, it was not entirely satisfactory.

For one thing, the gradual shrinking of the Sun was a hypothesis lacking observational confirmation. Thomson calculated that the Sun's diameter would decrease with 70 m per year, but careful studies made by astronomers showed no trace of the predicted diminution. For another thing, the theory predicted a lifetime 3of the Sun much smaller than the age of the Earth as advocated by geologists and natural historians. Whereas Thomson concluded that the Sun could not have existed as a strongly radiating body for more than 20 million years, and that it would cease to shine appreciably in another 5 or 6 million years, the geologists claimed that the Earth was at least 100 million years old and probably as old as 300 million years (Burchfield 1975, 134-139).

---

[1] Given Helmholtz's priority, from a historical point of view "Helmholtz-Thomson" seems a more appropriate term. In modern astrophysics the contraction theory is known as the KH (Kelvin-Helmholtz) mechanism. Although this mechanism does not apply to the Sun and other main-sequence stars, it does apply to certain other stellar and planetary objects.

To make a long story short, by the turn of the century the Thomson-Helmholtz theory had lost much of its former authority, and yet it lived on for a decade or more, the reason being that there was no credible alternative. As Arthur Eddington stated in an address to the British Association for the Advancement of Science: "Only the inertia of tradition keeps the contraction theory alive – or, rather, not alive, but an unburied corpse" (Eddington 1920, 18).

The American geologist Thomas C. Chamberlin was a sharp critic of the Thomson-Helmholtz theory. Without referring to the recent discovery of radioactivity, in a paper of 1899 he pointed out that the theory ignored other forms of energy such as those presumably hidden in the interior of atoms:

> What the internal constitution of the atoms may be is yet an open question. It is not improbable that they are complex organizations and the seats of enormous energies. Certainly, no competent chemist would affirm either that the atoms are really elementary or that there may not be locked up in them energies of the first order of magnitude. … Why should not atoms, atomecules[2], and whatever lies below, one after another have their energies squeezed out of them; and the outer regions be heated and lighted for an unknowable period at their expense? (Chamberlin 1899)

Chamberlin did not follow up on his prophetic speculation, but within a few years, subatomic energy entered the question of the Sun's emittance of light and heat. During the first decade of the twentieth century, many scientists considered radioactivity to be an attractive alternative to gravitational contraction (Hufbauer 1981; Kragh 2016).

In 1903 Pierre Curie and his assistant Pierre Laborde announced that they had determined the heat energy produced by radium decay to 100 calories per gram per hour or no less than about 200,000 times the energy released in the burning of coal. Might not the Sun be a radioactive machine, perhaps consisting of radium or some other highly radioactive substances? This is indeed what many scientists believed. Among those who supported the hypothesis, or at least found it to be promising, were notables such as Johannes Stark, Ernest Rutherford, George H. Darwin, Frederick Soddy, and Henri Poincaré. In his 1904 monograph *Radio-Activity*, Rutherford estimated that "the presence of radium in the sun, to the extent of 2.5 parts by weight in a million, would account for its present rate of emission of energy." He went along in a more speculative mood:

> It is not improbable that, at the enormous temperature of the sun, the breaking up of the elements into simpler forms may be taking place at a more rapid rate than on the earth. If the energy resident in the atoms of the elements is thus available, the time during which the sun may continue to emit heat at the present rate may be from 50 to 500 times longer than was computed by Lord Kelvin from dynamical data. (Rutherford 1904, 344)

---

[2] Chamberlin probably thought of the "atomecule" as a hypothetical subatomic particle, possibly with the electron in mind. The term was introduced by the British philosopher H.M. Stanley (1884), but Chamberlin may have been the only scientist to use it.

As late as 1913, in his monograph *Radio-Active Substances and Their Radiations*, Rutherford repeated the speculation.

For a period of time the hypothesis of a radioactive Sun was popular, but it was never developed into a proper scientific theory. Because of its qualitative nature, it was not possible to calculate the radioactive energy flux and compare it to the solar constant without arbitrary assumptions. Nor did the hypothesis fare well with observations. If the source of the Sun's heat were radioactive elements, one would expect their spectral lines to turn up in the spectra of the solar atmosphere, which they did not. At around 1910 the radioactive solar hypothesis was realised to be a blind alley, nothing but a speculation. In regard of the general disbelief in the traditional contraction theory it meant that the source of the Sun's energy was as mysterious as ever.

## 3. Eddington on stellar energy

Perhaps inspired by the earlier ideas of a radioactive Sun, in around 1920 a few scientists argued that stellar energy was of subatomic or nuclear origin. Essentially two very different processes were considered, both of them hypothetical and only one of which has survived to this day.[3] In a paper of 1917, Eddington tentatively suggested that under the conditions in the interior of stars matter might annihilate as a result of proton-electron annihilation, that is, $p^+ + e^- \rightarrow$ energy (Eddington 1917; Hufbauer 1981). He was not the only one to argue for this kind of process, which also figured prominently in the works of James Jeans from about 1924 to 1932 (e.g.

---

[3] For the sake of brevity, I skip a third nuclear process which played some role in the 1920s, namely the radioactive decay of hypothetical transuranic elements. The hypothesis was in different versions defended by Walther Nernst and James Jeans, but it was rejected by Eddington and the majority of physicists and astronomers (Kragh 2013).

Jeans 1926). In fact, Jeans had suggested annihilation between negative electrons and hypothetical positive electrons many years earlier, but it was only with Eddington that the idea, now as proton-electron annihilation, was applied to the problem of stellar energy (Douglas 1956, 68).

Three years later Eddington came up with the idea that stellar energy might be due to the building up of helium nuclei from protons and electrons, a process which can be written as

$$4p^+ + 2e^- \rightarrow \text{He}$$

At the time it was known from Francis Aston's experiments with the new mass spectrometer that the mass of a helium nucleus is about 0.7% less than that of four protons, meaning that the mass difference $\Delta m$ would be emitted as radiation energy in accordance with Einstein's formula $E = \Delta m c^2$. "If 5 per cent of a star's mass consists initially of hydrogen atoms," Eddington (1920) wrote, "the total heat liberated will more than suffice for our demands, and we need look no further for the source of a star's energy."

The idea of helium fusion from hydrogen had first been suggested by the American chemist William Harkins, who in 1917 suggested the 4H → He process as the first step in the formation of complex atoms. However, Harkins did not relate his suggestion to stellar processes. On the other hand, this is what the French physicist Jean Perrin did in a couple of papers of 1920-1921 (Wesemael 2009). According to Perrin, solar energy might have its source in helium synthesis. Although his ideas were roughly similar to Eddington's, they were not well known and made very little impact on the community of astronomers.

Without clearly preferring one of the two alternatives, annihilation and helium synthesis, Eddington compared them in a chapter in his influential monograph of 1926, *The Internal Constitution of the Stars*. As he fully realised, both alternatives were problematic. Among the physical difficulties facing the fusion hypothesis was that it required the extremely improbable collision of six particles, which would only be possible if they moved at enormous velocities corresponding to a temperature in the interior of stars far exceeding the 40 million degrees that astronomers at the time estimated for the Sun. On the other hand, it had the advantage that it tied together the generation of stellar energy with the problem of how the heavier elements were formed in the stars.

The annihilation hypothesis was even more problematic as it was purely speculative and unsupported by experiments. Nonetheless, both Jeans and Eddington took annihilation very seriously. Although Eddington vacillated between the two possibilities of subatomic energy production, by the late 1920s he leaned towards the annihilation hypothesis, which not only was far more energy-efficient but also promised a much longer lifetime of stars ($1.5 \times 10^{13}$ years) than the fusion alternative ($1.5 \times 10^{11}$ years). Only in the mid-1930s did he abandon proton-electron annihilation, and then without declaring it an impossible process such as claimed by the quantum physicists.

Eddington was undoubtedly a pioneer of what came to be known as nuclear astrophysics, but he was reluctant when it came to developing the ideas of stellar energy production which he only discussed in a qualitative and uncommitted way. Despite his arguments for the fusion process, he did not claim that the true source of stellar energy was to be found in the building up of elements from hydrogen. As he wrote, "it is not of any great consequence whether in this suggestion we have actually laid a finger on the true source of the heat. It is sufficient if the discussion opens our eyes to the wider possibilities" (Eddington 1920). As seen in retrospect, his paper of 1920 is "a milestone in the search for the energy source in stars" (Shaviv 2009, 148), but this is not how it was seen at the time and nor was it how Eddington perceived it.

Without attempting to harvest the fruits of the seeds he had planted, Eddington left the field to a new generation of nuclear physicists, apparently uninterested in their results. It is quite remarkable that he ignored the progress which in the 1930s finally led to an understanding of the mechanism of stellar energy and which in a qualitative sense rested upon his early ideas. Thus, nowhere in his publications from 1939 to 1946 did he mention the breakthrough theory of Bethe. It may have been his obsession with a fundamental theory of everything that caused him to withdraw from stellar energy research (Hufbauer 2006; Kragh 2017b).

## 4. The beginning of nuclear astrophysics

The theories of stellar energy proposed by Eddington, Jeans and Perrin were classical, making no use of quantum theory in either the old Bohr-Sommerfeld version or the new Heisenberg-Schrödinger version of 1925-1926. When quantum mechanics appeared, it was for a time uncertain if it also applied to the enigmatic atomic nucleus, but in 1928 the Russian physicist George Gamow demonstrated that the emission of alpha particles from the nucleus could be understood on the basis of Schrödinger's wave mechanics. His work was independently duplicated by Ronald Gurney and Edward Condon at Princeton University whose paper in *Nature* appeared about a month before Gamow's more detailed paper in *Zeitschrift für Physik*. However, contrary to Gamow the two Americans concluded that the theory did not apply to the inverse problem of an alpha particle entering the nucleus (Stuewer 1986).

Directly building on Gamow's theory, in the spring of 1929 Robert d'Escourt Atkinson and Fritz Houtermans, a Briton and a German, co-authored an innovative paper which marked the beginning of nuclear astrophysics based on quantum mechanics (Atkinson and Houtermans 1929).[4]

Applying the reverse Gamow theory to fusion processes in the interior of a star, Atkinson and Houtermans assumed an environment of temperature 40 million degrees and density of approximately $10 \text{ g/cm}^3$. Under these conditions they suggested after extensive calculations that the main source of stellar energy might be the process envisaged by Eddington, that is, a transformation of four protons and two electrons into a helium nucleus. However, contrary to the improbable six-particle collision they imagined a cyclic process taking place by the consecutive capture of the six particles by a light nucleus and the subsequent expulsion of a helium nucleus in the form of an alpha particle. Although the Atkinson-Houtermans theory was a promising beginning, the two authors realised that it was far from the final solution to the intertwined problems of energy production and element formation in the stars. They described the formation of heavy elements as a "complete mystery" and left open the possibility that the source of stellar energy might be the annihilation of matter considered by Eddington and Jeans.

At a time when nuclear astrophysics was foreign to the large majority of astronomers and only of marginal interest to most physicists, the paper by Atkinson and Houtermans attracted little immediate interest (according to the *Web of Science* database it was only cited six times in the years 1929-1931). It became more widely known in 1931, when Atkinson published a more comprehensive and ambitious version of the theory in *Astrophysical Journal* (Atkinson 1931a). In this paper, Atkinson took advantage of the recent recognition, only obtained about 1930, that hydrogen is by far the most abundant element in the Sun's atmosphere. Just five years earlier it was generally accepted that the stars have approximately the same chemical composition as the Earth, but in 1928-1930 works by Cecilia Payne (later Payne-Gaboschkin), Henry Norris Russell and others changed the picture completely. The new result suggested that hydrogen might also be the predominant element in the interior of the Sun and other stars, such as Atkinson assumed.

---

[4] Originally the two physicists titled their paper "Wie kann man ein Helium Kern in einen potential Topf köchen?" (How can one Cook a Helium Nucleus in a Potential Pot?), but the editor of *Zeitschrift für Physik*, the German physicist Karl Scheel, found it too imaginative for a respectable scientific journal (Gamow 1970, 73).

Atkinson not only dealt with stellar nuclear reactions but also speculated that his theory might be of cosmological significance. In the summer of 1931, shortly after the expanding universe had replaced the earlier static models, he suggested a scenario of how the present universe might have evolved from a primordial state:

> It thus appears that as a result of the wave mechanics on one hand, and the general theory of relativity on the other, the universe may have developed its present complexity of stars and of atoms from an initial state consisting of a fairly dense, nearly uniform, nearly stationary mass of cold hydrogen. This comparatively simple beginning constitutes at least a pleasant ornament, if not an actual support for our theory. (Atkinson 1931b).

The scenario published in the 1 August 1931 issue of *Nature* was one of the first ever of what later became known as the big-bang theory of the origin of the universe. Indeed, it had elements in common with Georges Lemaître's article in the same journal of 9 May 1931 in which the "primeval atom" or big-bang idea was first introduced. On the other hand, the starting points of the two cosmological scenarios were different: whereas Atkinson assumed a dense cloud of hydrogen atoms, Lemaître's world started in a pre-atomic state where all matter was condensed into a huge atomic nucleus. Contrary to Lemaître, Atkinson did not develop his idea, which may have been just a casual remark. All the same, it indicates that he was familiar with the expanding universe and relativistic cosmology generally.[5]

Although the Atkinson-Houtermans theory and its development into Atkinson's more comprehensive theory of 1931 was an important step forwards, its impact was limited and not generally regarded as a solution to the stellar energy puzzle. As an indication of the uncertainty it is worth recalling that at the time Niels Bohr seriously considered that the fundamental law of energy conservation might be invalid in the nuclear processes taking place in the interior of stars (Kragh 2017a). Bohr first discussed the heretical hypothesis of stellar energy non-conservation in 1929 and returned to it over the next three years if only as a vaguely formulated suggestion. Although few physicists and even fewer astronomers took it seriously, for a while young Lev Landau found it to be justified. In an important paper of 1932 dealing with stars at very high density, Landau referred approvingly to Bohr's speculation:

---

[5] "General relativity leads of course to the view that space as a whole, in which are imbedded the extra-galactic systems, is expanding, and the motion is also observed, as is well-known." Atkinson (1931a).

> Following a beautiful idea of Prof. Niels Bohr's we are able to believe that the stellar radiation is due simply to a violation of the law of energy, which law, as Bohr has first pointed out, is no longer valid in the relativistic quantum theory, when the laws of ordinary quantum mechanics break down … We expect that this must occur when the density of matter becomes so great that atomic nuclei come in close contact, forming one gigantic nucleus. (Landau 1932)

By the early 1930s nuclear processes in the interior of stars were not yet considered an important part of astrophysics, and the research area was generally foreign land to astronomers and classically trained astrophysicists. As seen from the perspective of most quantum physicists, nuclear astrophysics was a marginal subject that attracted interest only among a minority of physicists. The low status of the subject is illustrated by its absence in the eight volumes of the authoritative *Handbuch der Astrophysik* published between 1928 and 1933. Nuclear astrophysics only entered in a supplementary volume from 1936, where it was covered in a chapter written by the Danish astronomer Bengt Strömgren and then without giving it much space.

## 5. An annus mirabilis and what followed

Although the first cautious applications of nuclear physics to areas of astronomy date from the early 1920s, it was only with the annus mirabilis of 1932 that nuclear astrophysics took off as a promising interdisciplinary branch of science. What made the year miraculous from an astrophysical point of view was the discovery of new elementary particles (neutron, deuteron, positron), and generally the recognition that atomic nuclei consist of protons and neutrons. Until about 1932 it was universally believed that the nucleus was a composite of protons and electrons, which was the assumption underlying all theories from Eddington in 1917 to Atkinson in 1931. For example, the helium nucleus was assumed to consist of four protons bound together by two electrons, $^4\text{He} = (4p^+, 2e^-)$.

Two lines of interconnected research programs followed the annus mirabilis, the one focusing on the synthesis of elements in stars and the other on the nuclear processes responsible for their energy production. With the advent of high-energy accelerators, it proved possible for the first time to test in the laboratory the theorists' ideas of possible thermonuclear reactions.

The early phase on nuclear astrophysics was completely dominated by physicists, whereas astronomers and classically trained astrophysicists hesitated in entering or even acknowledging the field. Gamow, an expert in nuclear theory, was one of the physicists who turned to astrophysics, an area he considered to be just nuclear physics applied to the stars. He was particularly fascinated by the prospect of explaining the relative abundances of the chemical elements by means of nuclear processes, a problem he attacked by assuming consecutive neutron capture to be the basic mechanism. He first considered the possibility in an address of 1935, in which he also argued that the long-discarded Thomson-Helmholtz contraction theory might play a significant role in stellar energy production. Gamow believed at the time that stars contained a high-density core of tightly packed neutrons and that gravitational contraction to this kind of matter might provide the main source of stellar energy: "The question of the mechanism of energy-liberation is not yet quite clear but one can easily see that gravitational energy alone liberated in the contraction (Helmholtz's old theory) to such large densities would be enough to secure sufficiently long periods for stellar lives" (Gamow 1935).

Gamow's attempt to resuscitate the gravitational contraction theory for ordinary stars was unsuccessful and he soon returned to the safer haven of nuclear processes. In a work of 1938, he assumed an original stellar state of pure hydrogen from which deuterons (*d*) and neutrons (*n*) were produced by the processes

$$p + p \rightarrow d + e^+$$

and

$$d + d \rightarrow {}^3\text{He} + n,$$

where the symbol $e^+$ denotes a positron. Neutron captures and subsequent beta decays would then result in the building up of still heavier elements, or so he believed. The first process had previously been considered by Atkinson, who like Gamow argued that a mixture of deuterons, protons and neutrons might lead to a synthesis of helium and still heavier elements.

Atkinson, Gamow and other pioneers of nuclear astrophysics were aware of and to some extent inspired by contemporary accelerator experiments made by British and American physicists. For example, in 1934 Rutherford and his collaborators Mark Oliphant and Paul Harteck at the Cavendish Laboratory bombarded deuterium in the form of ammonium chloride $ND_4Cl$ with high-speed deuterons, reporting the reactions

$$d + d \rightarrow {}^3\text{He} + n$$

and

$$d + d \rightarrow {}^3\text{H} + p$$

The second of the reactions is today widely cited as the discovery of tritium, the hydrogen mass-3 isotope. As to the first reaction it appeared with a surplus energy of 3.2 MeV, but Rutherford and his collaborators did not associate the fusion process with the astrophysicists' ideas of stellar energy production.

The inspiration from and analogy to laboratory experiments was spelled out by the British physicist Harold Walke in an early paper on stellar nuclear synthesis:

> The atomic physicist with his sources of high potentials and his discharge tubes, is synthesizing elements in the same way as is occurring in stellar interiors, and the processes observed, which result in the liberation of such large amounts of energy of the order of millions of volts indicate how the intense radiation of stars is maintained and why their temperatures are so high. (Walke 1935)

Walke's own and ill-fated theory assumed neutrons and not protons to be the initial state of stellar matter. As an illustration of the uncertainty in nuclear and elementary particle physics still reigning at the time, he thought that the proton was a derivative of the neutron, namely a neutron-positron composite, $p = (n, e^+)$.

Another ill-fated theory was proposed by the Estonian astronomer Ernst Öpik, who in 1938 published a comprehensive review of stellar structure, evolution and energy in *Publications de l'Observatoire Astronomique de l'Université de Tartu*.[6] According to Öpik, energy generation in main-sequence stars was initiated by proton-proton synthesis to deuterons, the same kind of process that Gamow had postulated. However, Öpik realised that the process was purely hypothetical and, if it existed, might lead to an absurd scenario. "We are here confronted with a dilemma," he wrote. "If the reaction were detectable experimentally, the Sun would blow up from the immense energy generation; or rather, the Sun would exist only as a diffuse star … of about nine times its present radius" (Öpik 1938, 28). Might it be, he speculated in stark contrast to the consensus view, that the interior of the Sun was devoid of hydrogen? Having contemplated the question, he decided that a hydrogen-free solar core was after all not possible. Öpik also considered the experimentally verified lithium-hydrogen reaction

$$^7\text{Li} + p \rightarrow {}^4\text{He} + {}^4\text{He}$$

as a source of solar energy, but had to admit that the stars contained too little lithium for the reaction to be of any importance.

---

[6] The full text of Öpik's 115-page long essay on "Stellar Structure, Source of Energy, and Evolution" is accessible on http://articles.adsabs.harvard.edu/pdf/1938PTarO..30C...1O.

Whereas Öpik's work published in the proceedings of the Tartu Observatory was not widely circulated, the theory of element formation proposed by Carl Friedrich von Weizsäcker in two papers in *Physikalische Zeitschrift* was well known and influential (Shaviv 2009, 317-325; Hufbauer 2006, 219-220). In a paper of 1937, 25-year old Weizsäcker started out, like Atkinson and Gamow before him, with proton-proton reactions and proposed further reaction schemes from which elements heavier than helium could be formed. However, he ran into insuperable difficulties and consequently abandoned in the second paper of 1938 the assumption that the stars began their lives in a form of pure hydrogen. Motivated by discussions with Gamow and Strömgren, he now argued that element formation took place in a primordial state of the universe characterized by the extreme temperature of about $10^{11}$ degrees and a no less extreme density in the neighbourhood of the density of an atomic nucleus, of the order $10^{14}$ g/cm$^3$. Thus, he separated the origin of the elements from the energy of the stars. "It is quite possible," Weizsäcker (1938) wrote, "that the formation of the elements took place before the origin of the stars, in a state of the universe significantly different from today's."

Weizsäcker's cosmological approach had the advantage that he could make use of other elements to propose a new cyclic model for the generation of stellar nuclear energy (Wiescher 2018). He briefly considered other possibilities of energy production, such as gravitational contraction and mass annihilation, but only to reject them. What became known as the CN cycle (or CNO cycle), a model which shortly later would be presented in greater detail by Bethe, presumed the catalytic action of carbon-12 nuclei and also involved nuclei of nitrogen and oxygen. The net process was 4H → He, the same process which was originally contemplated by Harkins and Eddington. Although Weizsäcker was the first to state the CN cycle, he discussed it in qualitative terms only and did not, contrary to Bethe, calculate the energy release. The cycle is sometimes referred to as the Bethe-Weizsäcker process but only very rarely as the Weizsäcker-Bethe process.

In retrospect, Weizsäcker's 1938 article was more than just a contribution to the old problem of why the stars shine. It was also and even more importantly a pioneering attempt to provide the origin of the universe with a physical explanation. German physicists and astronomers widely interpreted his theory as implying a primeval explosion or what later would be called a big bang (Heckmann 1942, 100).

## 6. Conclusion: A Nobel Prize in astrophysics

"The question about nuclear reaction which gives rise to stellar energy seems to be finally settled." Thus wrote Gamow to the prominent American astronomer Henry

Norris Russell in a letter from the fall of 1938 (DeVorkin 2000, 253). He referred to a forthcoming monumental paper in *Physical Review* in which Bethe examined in great detail the CN cycle, arguing that it was responsible for the energy production of the Sun and other main-sequence stars. The previous year, in a joint paper with Charles Critchfield, he had investigated the proton-proton or *pp* cycle, which leads to the same net result (4H → He) but varies less drastically with the temperature than the CN cycle. Bethe calculated that the temperature dependence was $T^{3.5}$ for the *pp* cycle and $T^{16}$ for the CN cycle.[7]

Contrary to Weizsäcker, Bethe focused exclusively on the energy and refrained from speculating on a primeval state of stellar matter or the synthesis of heavier elements. "We can … drop the discussion of the building up of elements entirely and can confine ourselves to the energy production which is, in fact, the only observable process in stars" (Bethe 1939; Wiescher 2018). He also took into account the gamma quanta and neutrinos associated with the nuclear processes, which Weizsäcker did not. Bethe deduced that the CN cycle, in order to reproduce the energy emitted by the Sun, would require a central temperature of 18.5 million degrees. This theoretically derived temperature agreed impressively with the value 19 million degrees based on astrophysical models of the Sun.

Bethe's theory of stellar energy became an instant success, universally hailed as a breakthrough in nuclear astrophysics. According to Russell, praising the theory in *Scientific American*, it was "the most notable achievement of theoretical astrophysics of the last fifteen years" (DeVorkin 2000, 255). In 1967 Bethe was belatedly awarded the full Nobel Prize in physics for "his contributions to the theory of nuclear reactions, especially for his discoveries concerning energy production in stars."[8] Time was more than ripe for Bethe's prize, for until the year of his award he had been nominated no less than 48 times by a series of prominent physicists and astronomers, first in 1943 by Otto Struve at Yerkes Observatory. However, until 1967 the Nobel Committee hesitated in recognizing astrophysics as a proper branch of physics, such as the committee conceived the term (Kragh 2017c). In his presentation speech in Stockholm, the Swedish physicist Oskar Klein justified the physics prize to Bethe by arguing that the solution to the riddle of stellar energy qualified as fundamental physics. Incidentally, also Gamow and Weizsäcker were nominated for a Nobel Prize for their contributions to nuclear astrophysics, but in their cases with no success (Wiescher 2014).

---

[7] According to present knowledge, the core temperature of the Sun is 15.7 million degrees. It follows that the Sun is mainly powered by the *pp* cycle (about 90%) and not by the CN cycle (about 10%), as Bethe concluded in 1939.

[8] https://www.nobelprize.org/prizes/physics/1967/summary/